# Electron mobility in n-doped zinc sulphide

Clóves G. Rodrigues

*Núcleo de Pesquisa em Física, Departamento de Física, Universidade Católica de Goiás, CP 86, 74605-010 Goiânia, Goiás, Brazil*

**Abstract**

A study of the mobility of n-doped wurtzite and zincblende ZnS is reported. We have determined nonequilibrium thermodynamic state of the ZnS—driven far away from equilibrium by a strong electric field—in the steady state. The dependence of the mobility (which depends on the nonequilibrium thermodynamic state of the sample) on the electric field strength is derived, which decreases with the strength of electric field. We analyzed the contributions to the mobility arising out of the different channels of electron scattering, namely, the polar optic, deformation, interactions with the phonons, and with ionized impurities. The case of *n*-ZnS WZ and ZB have been analyzed: as expected the main contribution comes from the polar-optic interactions in this strong-polar semiconductor. The other interactions are in decreasing order, the deformation acoustic and the one due to impurities.

## 1. Introduction

There exists nowadays a large interest in the study of large-gap semiconductors, because of their use in devices, like diodes and lasers [1–4].

The zinc sulfide (ZnS) is a wide band-gap semiconductor, which crystallizes both in zincblende (ZB) and wurtzite (WZ) phases. The ZnS has been investigated in a host of new materials as an application of thin-film technology [5–7]. Thin-film electroluminescent devices have become of great interest since they offer a possible means of achieving a high-resolution, light-weight, compact video displays panel for computer terminals or television screens (flat-panel displays). The electroluminescent devices offer significant advantages over other existing technologies such as cathode ray tubes, plasma, and liquid-crystal displays [8,9].

From a device applications point of view, it is extremely important to compare the properties of the different phases of a particular material. For instance, one phase can be more suitable than another in some applications or one phase can be more easily grown making it more attractive for certain device applications. As a general rule, the ZnS transport properties have been calculated using Boltzmann transport equations [10] and Monte Carlo simulations [11,12]. Seeking a better understanding of the electron transport in zincblende and wurtzite ZnS, it was performed in this work a theoretical study of their transport properties resorting to a powerful, concise, and soundly based kinetic theory for far-from equilibrium systems. It is the one founded one on a nonequilibrium statistical ensemble formalism, the so-called MaxEnt-NESOM for short [13–15], which provides an elegant, practical, and physically clear picture for describing irreversible processes [16], as for example in semiconductors far-from equilibrium [17,18], which is the case considered here. Through the numerical solution of associate quantum transport equations based on the MaxEnt-NESOM, it was characterized the electron mobility in WZ and ZB ZnS, doped type *n*.

## 2. Mobility of electrons

Consider the case of n-doped direct-gap polar semiconductors, in contact and in equilibrium with a reservoir at temperature $T_0$. An electric field $\mathbf{F}$ is applied, in, say, $x$ direction. It drives the system out of equilibrium, and the time-dependent (due to the relaxation processes that unfold in it) macroscopic state is described in terms of a statistical thermodynamics for irreversible systems, namely Informational Statistical Thermodynamics [16], which is based on the so-called Predictive Statistical Mechanics [15].

In the present case, the nonequilibrium thermodynamic state is characterized by the macrovariables carriers' energy, $E_e(t)$, carriers' linear momentum, $\mathbf{P}(t)$, along the $x$-axis, and the energies of the longitudinal optical and acoustic phonons, $E_{LO}(t)$ and $E_{AC}(t)$, respectively (all are given per unit volume). The TO phonons have been ignored once it is negligible the deformation potential interaction with electrons in the conduction band [19]. The equations of evolution for these basic macrovariables are derived in the nonlinear quantum kinetic theory described in Ref. [20]. They are

$$\frac{d}{dt}N(t) = 0, \tag{1}$$

$$\frac{d}{dt}E_e(t) = -\frac{e\mathbf{F}}{m^*}\bullet\mathbf{P}(t) + J^{(2)}_{E_e}(t), \tag{2}$$

$$\frac{d}{dt}\mathbf{P}(t) = -ne\mathbf{F} + \mathbf{J}^{(2)}_{\mathbf{P}}(t), \tag{3}$$

$$\frac{d}{dt}E_{LO}(t) = J^{(2)}_{LO}(t) - J^{(2)}_{LO,an}(t), \tag{4}$$

$$\frac{d}{dt}E_{AC}(t) = J^{(2)}_{AC}(t) + J^{(2)}_{LO,an}(t) - J^{(2)}_{AC,dif}(t). \tag{5}$$

where $E_e$ is the electrons' energy and $\mathbf{P}$ their linear momentum; $E_{LO}$ is the energy of the longitudinal optical phonons, which strongly interact with the carriers via Fröhlich potential in this strong-polar semiconductor; $E_{AC}$ is the energy of the acoustic phonons, which play a role of a thermal bath; and $\mathbf{F}$ stands for the constant electric field in the $x$-direction. Let us analyze these equations term by term. Eq. (1) accounts for the fact that the concentration $n$ of electrons is fixed by doping. In Eq. (2) the first term on the right accounts for the rate of energy transferred from the electric field to the carriers, and the second term accounts for the transfer of the excess energy of the carriers - received in the first term—to the phonons. In Eq. (3) the first term on the right is the driving force generated by the presence of the electric field. The second term is the rate of momentum transfer due to interaction with the phonons and ionized impurities. In Eqs. (4) and (5) the first term on the right describes the rate of change of the energy of the phonons due to interaction with the electrons. More precisely they account for the gain of the energy transferred to then from the hot carriers and then the sum of contributions $J^{(2)}_{LO}(t)$and $J^{(2)}_{AC}(t)$is equal to the last term in Eq. (2), with change of sign. The second term in Eq. (4) accounts for the rate of transfer of energy from the optical phonons to the acoustic phonons via anharmonic interaction. The contribution $J^{(2)}_{LO,an}(t)$is the same but with different sign in Eqs. (4) and (5). Finally, the diffusion of heat from the AC phonons to the reservoir is account for in the last term in Eq. (5). The detailed expressions for the collision operators are given in Ref. [20].

We notice that the linear momentum density can by related to the drift velocity $\mathbf{v}(t)$ by the relation

$$\mathbf{P}(t) = Nm^*\mathbf{v}(t), \tag{6}$$

where $m^*$ is the effective mass of the electron, related to the current $\mathbf{I}$ by the expression

$$\mathbf{I}(t) = -ne\,\mathbf{v}(t), \tag{7}$$

which flows in the direction of the electric field. Moreover we define a momentum relaxation time, associated to the scattering of carriers by phonons and ionized impurities, given by

$$\tau_{\mathbf{P}}(t) = n\frac{m^* v(t)}{J^{(2)}_{\mathbf{P}}(t)}, \tag{8}$$

where, once we take into account that the collision operator $\mathbf{J}^{(2)}_{\mathbf{P}}(t)$is composed of the three contributions consisting of scattering by: optical (or Fröhlich) interaction with LO phonons ($=\mathbf{J}^{(2)}_{\mathbf{P},LO}(t)$), deformation potential with AC phonons ($=\mathbf{J}^{(2)}_{\mathbf{P},DA}(t)$), and scattering by ionized impurities ($=\mathbf{J}^{(2)}_{\mathbf{P},imp}(t)$), that is, $\mathbf{J}^{(2)}_{\mathbf{P}}(t) = \mathbf{J}^{(2)}_{\mathbf{P},LO}(t) + \mathbf{J}^{(2)}_{\mathbf{P},DA}(t) + \mathbf{J}^{(2)}_{\mathbf{P},imp}(t)$. We have then a Mathiessen-like rule of the form

$$\frac{1}{\tau_{\mathbf{P}}} = \frac{1}{\tau_{PO}} + \frac{1}{\tau_{DA}} + \frac{1}{\tau_{imp}}. \tag{9}$$

In Eq. (9), PO refers to polar-optic interaction, DA to acoustic deformation potential and imp to the effect of ionized impurities. We have introduced the individual relaxation times $\tau_j$

$$\tau_j = \frac{nm^* v(t)}{J^{(2)}_{\mathbf{P},j}(t)}, \tag{10}$$

where $J^{(2)}_{\mathbf{P},j}(t)$ is the contribution for $j$=PO, DA, and imp. Consequently, the mobility in the steady state, namely,

$$\mu = \frac{v}{F} = \frac{e}{m^*}\tau_{\mathbf{P}}, \tag{11}$$

(once, according to Eq. (3), in the steady state, $\mathbf{J}^{(2)}_{\mathbf{P}}(t) = neF$) becomes a composition of three contributions, according to the rule

$$\frac{1}{\mu} = \frac{1}{\mu_{PO}} + \frac{1}{\mu_{DA}} + \frac{1}{\mu_{imp}}, \tag{12}$$

## 3. Results

The set of coupled differential Eqs. (2)–(5) are solve numerically to obtain the steady-state behavior of the basic intensive nonequilibrium thermodynamic variables for zincblende and wurtzite $n$-ZnS. The Table 1 gives the parameters of wurtzite and zincblende phases that are used in these calculations. The doping concentration and the concentration of carriers are taken as $1\times10^{17}$ cm$^{-3}$, and the reservoir temperature 300 K ($T_0$=300 K). Numerical results are shown in Figs. 1 and 2.

Table 1
Parameters of ZnS

| Parameter | ZB | WZ |
|---|---|---|
| Electron effective mass $m_e^*$, $(m_0)$[a] | 0.34 | 0.28 |
| Lattice parameter $a$, (Å)[a] | 5.41 | 3.81 |
| Lattice parameter $c$, (Å)[a] | – | 6.26 |
| LO phonon energy $\hbar\omega_{lo}$, (meV)[a] | 43.0 | 42.6 |
| Static dielectric constant $\varepsilon_0$[a] | 8.32 | 9.6 |
| Optical dielectric constant $\varepsilon_\infty$[a] | 5.15 | 5.7 |
| Mass density ρ, (g/cm$^3$)[a] | 4.07 | 4.07 |
| Sound velocity, ($\times 10^5$ cm/s)[a] | 5.2 | 5.87 |
| Acoustic deformation potential $E_1$, (eV)[b] | 4.9 | 4.9 |

[a] Ref. [21].
[b] Ref. [22].

Fig. 1 shows the electric field dependence of the electron mobility in wurtzite (solid line) and zincblende (dashed line) phases. It can be noticed that the mobility decreases with increasing strength of the electric field, and that the larger electron mobility correspond to ZnS (WZ), $\mu^{(WZ)} \simeq 1.38\mu^{(ZB)}$, which can be ascribed to the electrons having a smaller effective mass in ZnS (WZ) than in ZnS (ZB) (see Table 1).

For the both phases, the Fig. 2 shows a comparison of the three contributions to the mobility (cf. Eq. (12)). It can be noticed that the main scattering mechanism is the polar optical-phonon scattering (Fröhlich interaction). Notice that is valid Mathiessen rule, and then the much smaller practically determines the value of the whole mobility. This is clearly a consequence that polar-optic (Fröhlich) interaction is strong in this compounds, and therefore produces a very short relaxation time as compared with the other scattering mechanisms (deformation acoustic and ionized impurities). Also it can be noticed that in the conditions of the calculations—$T_0 = 300$ K and $n = 10^{17}$ cm$^{-3}$—the scattering by the ionized impurities is negligible when compared to the others.

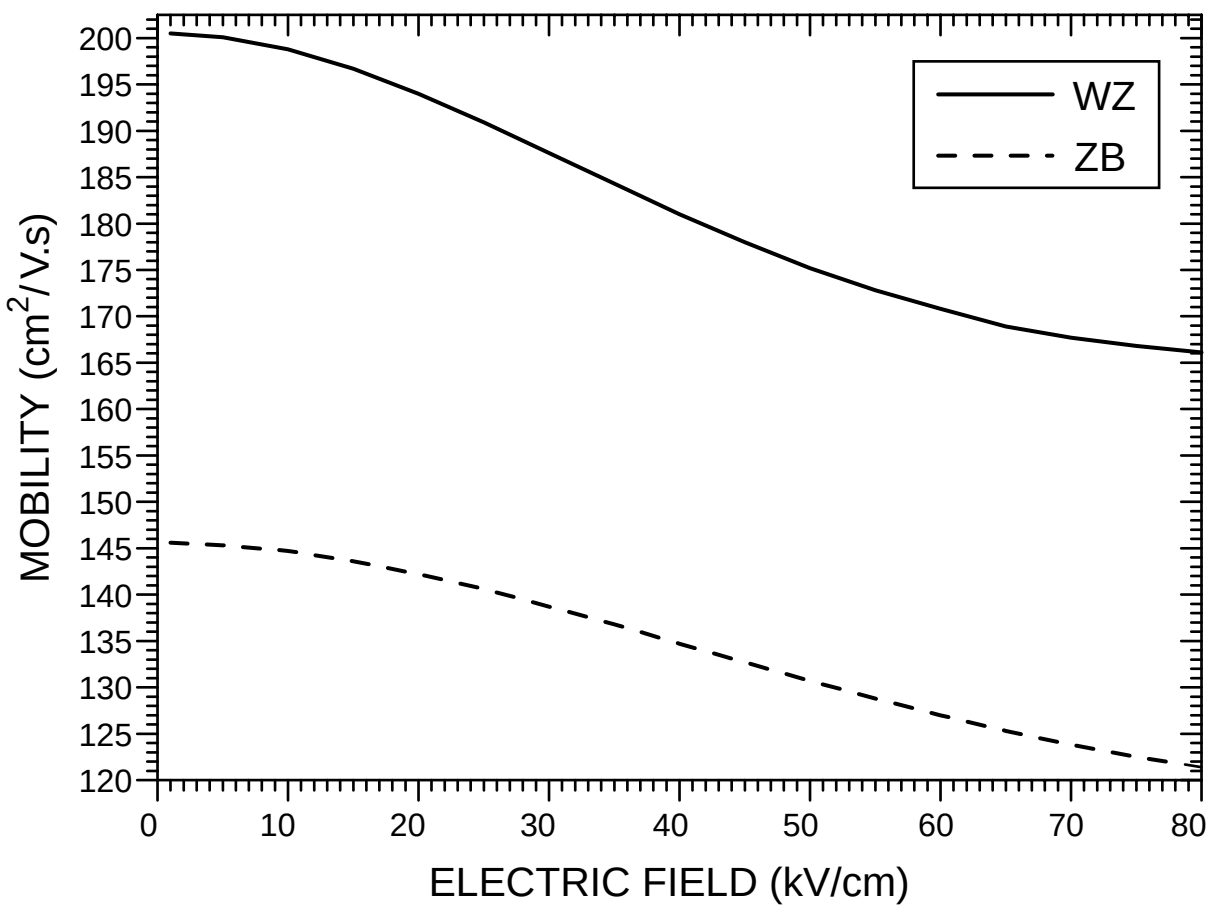

Fig. 1. Electron mobility as function of the electric field in ZnS: WZ phase (solid line), ZB phase (dashed line).

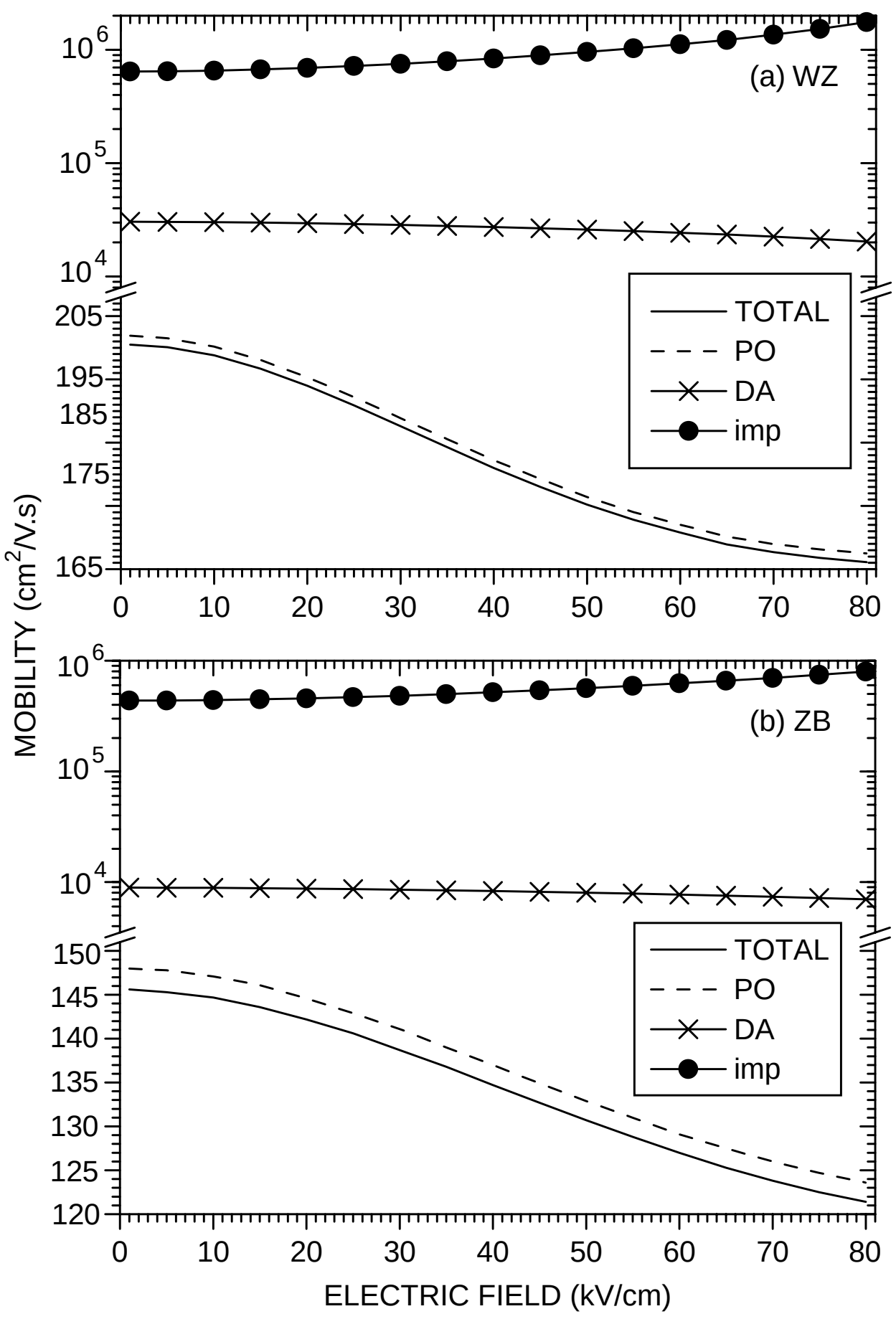

Fig. 2. Electron mobility as function of the electric field in ZnS for the different contributions; (a) WZ phase and (b) ZB phase.

## 4. Final comments

Transport properties were investigated, and we have concentrated our attention on these properties when the doped semiconductor is in the presence of moderated to high electric fields. In particular we report here a study of the mobility of ZnS (in wurtzite and zincblende phases) in such conditions. For that purpose we have determined the nonequilibrium thermodynamic state of the system—driven far-away from equilibrium by the electric field—in the steady state. The drift velocity (and hence the current) is derived—which of course depends on the nonequilibrium macroscopic (thermodynamic) state of the sample. The dependence of the mobility with the electric field strength is obtained, which decreases with increasing strength of the electric field. We have also analyzed the contributions to the mobility arising out of the different channels of electron scattering, namely, the polar optic, deformation, interactions with the phonons and with ionized impurities; at the room temperature, the main contribution comes from the polar-optic interactions in this strong-polar semiconductor. The other interactions are in decreasing order, the deformation acoustic and the ionized impurities. The deformation acoustic contribution is roughly 0.7% of the polar optic in wurtzite phase and 1.6% in zincblende phase.